\documentclass[10pt,twocolumn,letterpaper]{article}

\usepackage{cvpr}              

\usepackage{graphicx}
\usepackage{amsmath}
\usepackage{amssymb}
\usepackage{booktabs}

\usepackage[capitalize]{cleveref}
\crefname{section}{Sec.}{Secs.}
\Crefname{section}{Section}{Sections}
\Crefname{table}{Table}{Tables}
\crefname{table}{Tab.}{Tabs.}

\def\cvprPaperID{*****} 
\def\confName{CVPR}
\def\confYear{2022}

\begin{document}

\title{U-Net and its variants for Medical Image Segmentation : A short review}

\author{\textbf{Vinay} Ummadi\\
SMST, IIT Kharagpur\\
{\tt\small ummadi.vinay2000@gmail.com}
}

\maketitle

\begin{abstract}
  The paper is a short review of medical image segmentation using U-Net and its variants. As we understand going through a medical images is not an easy job for any clinician either radiologist or pathologist. Analysing medical images is the only way to perform non-invasive diagnosis. Segmenting out the regions of interest has significant importance in medical images and is key for diagnosis. This paper also gives a  bird eye view of how medical image segmentation has evolved. Also discusses challenge's and success of the deep neural architectures. Following how different hybrid architectures have built upon strong techniques from visual recognition tasks. In the end we will see current challenges and future directions for medical image segmentation(MIS).
\end{abstract}

\section{Introduction}
\label{sec:intro}

\begin{figure}
    \centering
    \includegraphics[scale=0.5]{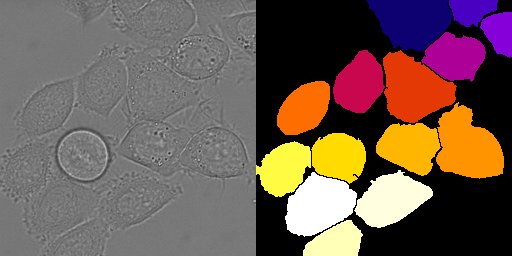}
    \caption{\textbf{Left} : Hela DIC image; \textbf{Right} : Instance cell segmentation}
    \label{fig:dic-cell}
\end{figure}

Let us first understand why do we need to do segmentation in medical images. Looking at images for disease diagnosis has been practiced since past few decades. Till date there are more than a dozen medical image modalities that used in the medical practice for \eg X-ray, Magnetic Resonance Imaging(MRI), Computer Tomography(CT), Ultrasound etc. Each of the image modality is selected based on speed of acquisition, image resolution, comfort to the patient. Once a image is acquired, a clinician will carefully inspect the image and will try to find the disease and possible causes. This process typically takes anywhere between few hours to few days depending on the problem complexity and also involves expert clinician, technicians in the loop. For example a pathologist will look at a large digital pathology image, identifies the different types of tissue regions, and look for abnormal regions which are his regions of interest. A radiologist may check the sizes of the different organs to check whether they normal or to be treated. A ophthalmologist may want to look for blood clots, or lipids in your retinal images. All of these tasks will require identifying regions of interest for \eg ophthalmologist is interested in finding blood vessel leakages. Segmentation is an implicit task performed by clinicians without mentioning it. That is why medical image segmentation(MIS) is significantly helpful for clinicians. An example of differential interference contrast microscopy image and cells segmented is shown in \cref{fig:dic-cell}. Medical practitioners want to track cells to understand their development and so they want to identify and segment the cells which is a segmentation and detection task.

How about automating this tedious task by a machine and final decision will be taken by a clinician. Image segmentation in medical images have been here from past few decades as early as 1980's. To the early 2014 image segmentation is performed though various methodologies, some simple methods are histogram threshold based segmentation, clustering based segmentation, mean shift segmentation, graph cut segmentation . Each of the simpler segmentation methods have their own advantages and disadvantages. But their application to medical image segmentation is limited. The best thing about these methods is they do not require any training that implies they do not depend on a data. In most cases these generalized segmentation algorithms would not work for medical image segmentation. Even when fine tuned to task at hand there is hardly any acceptable segmentation results. 

Deep learning started to gain traction starting 2010 and shown significant improvements in image classification tasks and outperformed existing state of the art. This success is mainly attributed to availability of large datasets and computational power of GPUs. Starting 2012 the convolutional neural networks outperformed all the existing traditional image processing methods on a variety of tasks including classification, detection, segmentation etc. General tasks in computer vision has reached reasonable performance on most datasets but this is not the case with the medical vision tasks. This is mainly due to data scarcity, data bias, cost of creating a labeled dataset. For example a model trained on patient data from hospital in USA to detect detecting diabetic retinopathy may work excellently well for those patient base and performs very poorly when tested at an Indian hospital. Many things will change from hospital to hospital for \eg local patient characteristics, image acquisition device, testing computer configurations etc. One of the main reason is data bias in the training set. Data augmentation is performed to increase the dataset size and which help in learning in variances which is an essential step unless we have a large dataset. Data prepossessing is one step before data augmentation which may involve intensity transformations, normalization and resampling.


Given the importance of medical image segmentation for clinician's and faster diagnosis, this area is under active research contributed by technical institutions, medical institutions and research labs across the globe. 



A simple classification scheme is chosen to introduce a few popular U-Net based deep neural networks for MIS. We start with a vanilla U-Net~\cite{unet} and head on to skip connections, residuals, recurrent, attention mechanism and transformer variants of U-net. U-Net is the one of the most popular architectures for different medical image segmentation tasks. Since the invention of U-Net many variants have built upon that. U-Net++ ~\cite{unet++} has more nested skip connections which helps in semantic learning and smooth gradient flow. R2U-Net~\cite{recurrentunet} is U-Net with both residual and recurrent connections. In the next section we will look at how these methods work and why they work for MIS. Attention U-Net~\cite{attunet} uses attention gates for improving the attention to regions of interest. Trans U-Net~\cite{transunet} is a hybrid network of U-Net and visual transformer that has attention modules. 

\section{Methodologies}
In this section, we present the brief ideas of the chosen methods and later look into their evolution and why each on them works well for some segmentation tasks.
%
\subsection{Old school segmentation}
In this section we will briefly look at some traditional image segmentation approaches that found applications in medical image segmentation. Threshold based segmentation involves finding a suitable threshold from histogram features and binarizing the image with that threshold. This method is not suitable when dealing multi-class segmentation and also cannot handle high degree intensity variations. Clustering based segmentation involves grouping similar pixels into clusters and assigning a color label for each cluster. K-means clustering is an typical clustering algorithm which segments the image into k clusters. The main disadvantage in method is number of cluster should be known beforehand. Mean shift segmentation is essentially a hill climbing algorithm which uses a window(w) and mean of the window is used to climb the hill. All pixels belonging to a hill are assigned to the same cluster. Graph Cut segmentation is most complex segmentation algorithm based on maximum flow and minimum cut algorithm. All of the traditional segmentation algorithms do a decent segmentation on natural images but they have shown poor performance in medical image segmentation. From we here we will briefly look at neural network based medical image segmentation methods.

\subsection{U-Net 2015}
\begin{figure}
    \centering
    \includegraphics[scale=0.5]{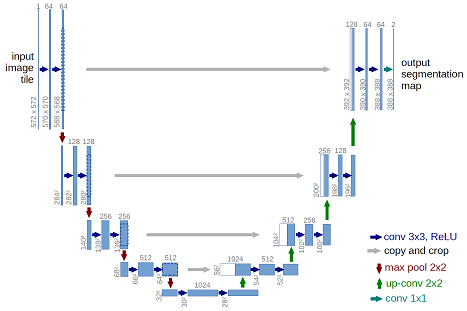}
    \caption{U-Net architecture}
    \label{fig:unet-arch}
\end{figure}

U-Net\cite{unet} (in \cref{fig:unet-arch}) is convolutional neural network proposed by \textit{Ronneberger}. It builds upon encoder-decoder architecture which is simply a hierarchical down-sampling convolutional layers followed by symmetric up-sampling convolutional layers, additionally feature maps from encoder network are concatenated in the respective decoder part for passage of semantic information. The left part is contracting path and called Encoder because it encodes information in higher spatial dimension to low dimensional latent representation. The right half part is expanding path called Decoder and it decodes the lower dimensional latent representation to higher dimensional segmentation map. The encoder contains convolutional(conv) layers with $3 X 3$ filters followed by a linear activation function ReLU(Rectified Linear Unit). Again a conv layer followed by ReLU. This conv-relu-conv-relu is called a conv block. A sub-sampling layer called Max Pool is used to reduce the spatial dimension of the feature map. The number of feature maps are progressively increased in the encoder. Decoder is a symmetric network of encoder, but instead of max-pool it uses up-convolution or transpose convolution to increase the spatial dimension of the feature map. Number of feature maps are reduces progressively throughout the decoder. Additionally features from encoder are cropped to fit with decoder feature map dimensions and concatenated. A Softmax layer is used for pixel wise classification which results in a segmentation map. U-Net is an end-to-end network trained with cross-entropy loss.

The reason U-Net being so popular for medical image segmentation, Often in medical tasks the availability of data is scarce and a model that learns with small quantity of data is in the need of hour. Surprisingly the exact requirement is fulfilled by U-Net, It outperformed other methods even with a very small number of images. When U-Net is trained on 35 partially annotated phase contrast light microscopy images it achieved an IoU(Intersection over Union) score of 0.92 which is best in 2015. The segmentation results are shown in supplementary figures material in Figure {1 and 2} in page 6.

\subsection{U-Net++ 2019}
\begin{figure}
    \centering
    \includegraphics[scale=0.2]{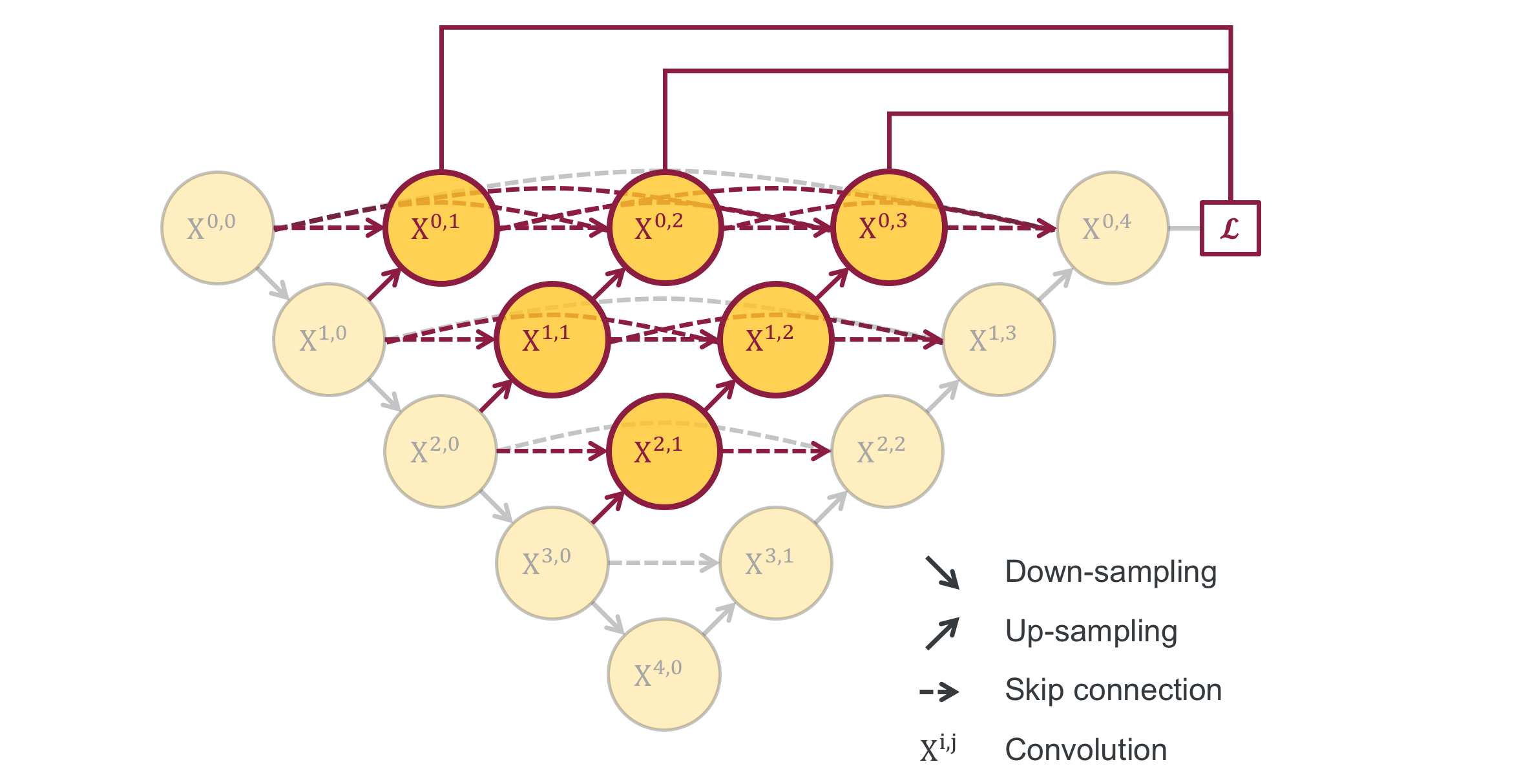}
    \caption{U-Net++ architecture}
    \label{fig:unetpparch}
\end{figure}

U-Net++ \cite{unet++} (in \cref{fig:unetpparch}) is a nested U-Net architecture that is densely connected \cite{densely2017} and deeply supervised. Precise segmentation is a strict requirement in MIS otherwise leads to poor performance in following interventional tasks. Nested and dense skip connections are used to supply right semantic and coarse graded information to the decoder part. These dense connections for convolutions are initially introduced by Huang \cite{densely2017} which have forward connections from all previous layers to current. These dense connections the help in vanishing gradient problem, feature reuse, strengthening feature propagation. Authors introduced deep supervision that helps in learning coarse -fine grained features at different levels. Segmentation maps are collected at final layers of sub networks and up-sampled to input scale for deep supervision. The main difference from U-Net are 
\begin{enumerate}
    \item Convolutional layers in the skip pathways
    \item Dense skip connections for gradient flow
    \item Deep supervision
\end{enumerate}
U-Net++ gains an average IoU score of 3.9 and 3.4 points on U-Net and wide U-Net respectively which is beneficial in many cases. The segmentation results are shown in supplementary figures material in Figure 3 in page 7 .

\subsection{R2U-Net 2018}
R2U-Net\cite{recurrentunet}(in \cref{fig:attnunet}) is another U-Net variant that builds upon popular residual and recurrent techniques. Deep convolutional networks perform well on most computer vision tasks. But there is limit for stacking layers which is due gradient propagation problems called vanishing gradient and exploding gradient. To solve the vanishing gradient, skip/identity connections are introduced. This is simply defined as identity connection from previous block to next block by which skipping current block operations. This work done \cite{resnet} Microsoft research Asia introduced a completely new paradigm of building very deep networks. Since then ResNets are very much used in all sorts of deep networks. Recurrent networks have been here since past three decades but their real application were only developed in the past decade. Recurrent networks and their variants are mostly used for sequential data \eg natural language, speech signals etc. Recurrent networks have recurrence feedback connections from output to input. This recurrence connection will eventually help in learning sequential context. R2U-Net is built upon levering recurrent and residual techniques. Each convolutional block in U-Net is replaced with a Recurrent-Residual block. Instead of crop and copy used in vanilla U-Net, a much simpler feature concatenation from encoder to decoder is used in R2U-Net. While having the similar number of parameters, R2U-Net shows a consistent performance improvement in empirical experiments. Authors have experimented on blood vessel segmentation, skin lesion segmentation and lung lesion segmentation where R2U shown to perform better than standard U-Net.
It is a common practice to evaluate segmentation with Sørensen–Dice coefficient and Jaccard score which are given by ${\displaystyle DSC={\frac {2|X\cap Y|}{|X|+|Y|}}}$ and ${\displaystyle JS ={{|X\cap Y|} \over {|X|+|Y|}}}$ respectively. On skin cancer lesion segmentation, R2U-Net has achieved a dice score of 0.86 and a standard U-Net achieved 0.84. In retinal blood vessel segmentation and lung lesion segmentation R2U-Net performed slightly better than a standard U-Net. The segmentation results are shown in supplementary figures material in Figure {4, 5, 6, 7, 8, 9} in page {8, 9, 10}.

\begin{figure}
    \centering
    \includegraphics[scale=0.25]{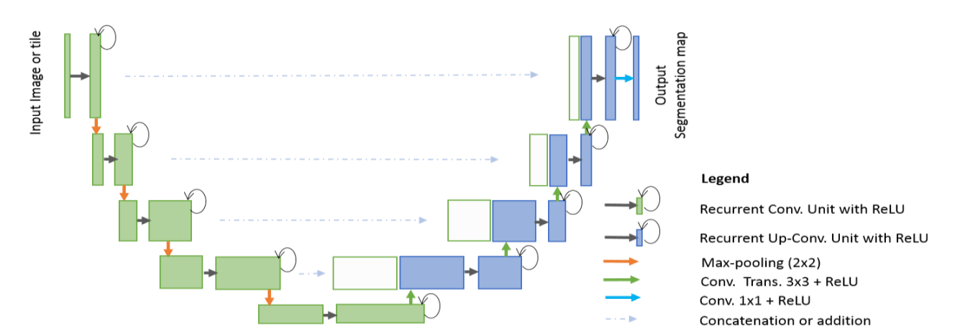}
    \caption{Recurrent Residual U-Net architecture} 
    \label{fig:r2unet}
\end{figure}
\subsection{Attention U-Net 2018}
\begin{figure}
    \centering
    \includegraphics[scale=0.175]{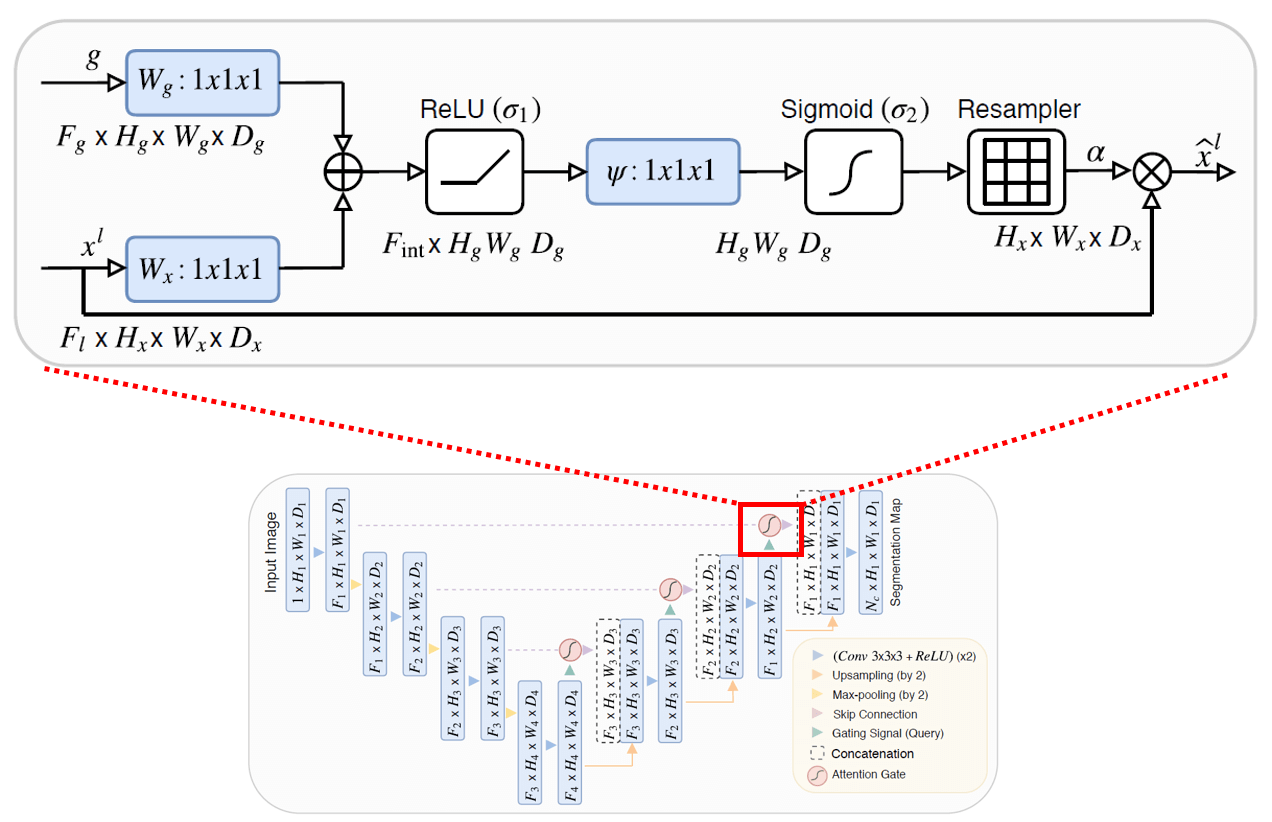}
    \caption{Attention U-Net architecture}
    \label{fig:attnunet}
\end{figure}

Attention U-Net\cite{attunet} (in \cref{fig:attnunet}) is a hybrid architecture with attention gates in the skip pathways. These attention gates will only pass salient features to the decoder and suppress redundant information for precise reconstruction of segmentation maps. Attention mechanism is at first applied for language tasks which shown tremendous success and idea is borrowed into visual tasks. Fundamentally attention gates help in learning the local context. There are two types of attention gates one is soft attention and another is hard attention. Soft attention outputs a weighted combination of inputs with weights between between 0 and 1 and this is a differentiable function. Attention U-Net uses soft attention gates in the skip pathways. Previously discussed models will suffer if the dataset has a large variations in the shape and size of the RoI. Attention gates implicitly highlight salient features and focus on RoI. Residual and dense connections are also present between encoder and decoder, which has their own advantages as mentioned in previous methods. Attention gates bring additional computational overhead, but increases the segmentation accuracy. Attention U-Net achieved a dice score of 81.48 ± 6.23 on CT82 dataset for pancreas segmentation. Attention U-Net consistently outperformed vanilla U-Net by a slim margin over a wide variety of tasks. The segmentation results are shown in supplementary figures material in Figure 10 in page 11 .

\subsection{Trans U-Net 2021}
\begin{figure}
    \centering
    \includegraphics[scale=0.35]{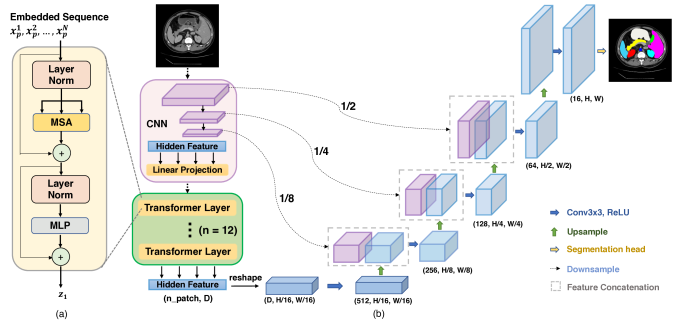}
    \caption{Transformer U-Net architecture}
    \label{fig:transunet}
\end{figure}

Fully convolutional networks(FCNs) show weak performance when there is a large variations in shape and size. Trans U-Net\cite{transunet} (in \cref{fig:transunet}) is an architecture that builds upon on the visual transformers(ViT).  Transformer networks have shown significant improvements for sequence to sequence prediction tasks like language translation and speech translation. Transformer contains multi head self-attention modules and that provide global self attention. Convolution neural networks demonstrate intrinsic locality. Since convolutions are local operations, U-Net fails to learn global spatial dependencies. Only by using transformers is also not a viable option since low level details are missing. Authors shown that by combining the U-Net and Transformer, we can achieve merits of both networks. Essentially TransUNet is combination of Transformer and U-Net. In this work Transformer encodes the patch wise image features with global representations, while a decoder upsamples the features to original image dimensions. Normal U-Net encoder is replaced by CNN based feature extractor and followed by transformer layers. While passing before to the decoder encoded CNN features and transformer features are concatenated appropriately. This encoding method helps in precise localisation of the RoI. Features are aggregated at multiple levels via skip connections. Under-segmentation and over-segmentation are reduced and better global contexts, better semantic information is learnt. Trans U-Net is evaluated on 30 CT abdominal scans from MICCAI abdomen labelling challenge and achieved an average DSC of 77.48 which is better than standard U-Net at 74.68. Authors also shown that Trans U-Net consistently outperformed even the best Attention U-Net. The segmentation results are shown in supplementary figures material in Figure 11 in page 11 .

\section{Discussion}
To once again recall that our main goal is to develop a precise medical image segmentation model. The objective of any MIS model is to deliver robust segmentation capabilities across majority  of medical image modalities. There is a need for universal robust MIS method.

The traditional segmentation algorithms fail to generalize for task specific MIS. The main advantage of these algorithms is that, they not need any training data at all. Before the development of U-Net, Fully convolutional networks(FCN) have without special attention to network design have partial success compared to traditional methods. MIS started to gain traction with only after U-Net, which revolutionized the field. The main contributions of the U-Net are 
\begin{enumerate}
    \item Generalized image segmentation architecture, can be used for arbitrary segmentation task
    \item High accuracy given proper data and training time
    \item Faster compared to other approaches
    \item Successful even with very limited training data
    \item Specifically successful in biomedical segmentation applications
\end{enumerate}
 
 The only limitation of the U-Net is it uses entire image, So when image is large a high GPU memory is required. 
 
 U-Net++ and R2U-Net has utilized the strong techniques like residual and dense connections that have worked on other visual tasks. While these methods slightly improve the performance but they are much complex. Later that Attention U-Net has also shown a slight improvement but it comes with high complexity and addition computational overheads. Only TransU-Net has a reasonable trade off for performance improvement and complexity of the transformer network. TransU-Net also requires more training data compared to U-Net to show reasonable improvements. While seen all four different variants of U-Net, expect TransU-Net other three has only marginal performance improvements. From 2016 to 2020 the progress in MIS is dull, only the arrival of transformers had put it again on the track.



    
To some extent medical image segmentation is reasonably accepted from both clinicians and engineers, still there are quite a few challenges exist. They are
\begin{enumerate}
    \item There are wide variety of medical images to perform segmentation.
    \item Limited availability of training data.
    \item Noisy labels ; Annotation bias between different clinical image annotators.
    \item Missing feedback loop between clinical experts and machine learning researchers.
\end{enumerate}

It is quite observable that neural architecture is to key to success in MIS. So Automatic neural architecture search for MIS can be a potential direction for researchers to find even better ones. Also known that every neural architecture is trained with a set of hyperparameters which led to success of the model. Automatic machine learning(AutoML) which can search for right neural architecture, data augmentation techniques, hyperparameters, loss functions would of great interest to the researchers. Model interpretation is in the line of interest to better understand models and build upon them. Apart from supervised learning approaches, self-supervised learning and weakly-supervised learning in medical imaging would also be potential directions to deal with weakly labeled and unlabeled data.

Finally want to thank my Professor Debashis Sen for teaching me Neural Networks and Computer Vision.


{\small
\bibliographystyle{ieee_fullname}
\bibliography{egbib}
}


\end{document}